\documentclass[conference]{IEEEtran}
\IEEEoverridecommandlockouts
\usepackage{cite}
\usepackage{amsmath,amssymb,amsfonts}
\usepackage{algorithmic}
\usepackage{graphicx}
\usepackage{textcomp}
\usepackage{xcolor}
\usepackage{hyperref}

\def\BibTeX{{\rm B\kern-.05em{\sc i\kern-.025em b}\kern-.08em
    T\kern-.1667em\lower.7ex\hbox{E}\kern-.125emX}}
    
\begin{document}

\title{Privacy-Preserving Spam Filtering\\ using Functional Encryption}

\author{
	\IEEEauthorblockN{
		Sicong Wang,\IEEEauthorrefmark{1}
		Naveen Karunanayake,\IEEEauthorrefmark{2}
		Tham Nguyen,\IEEEauthorrefmark{1}
		Suranga Seneviratne\IEEEauthorrefmark{1}
	}\\
	\IEEEauthorblockA{
		\IEEEauthorrefmark{1} University of Sydney, Australia, 
		\IEEEauthorrefmark{2} University of Moratuwa, Sri Lanka, \\
		Email: \{first name.last name\}@sydney.edu.au
	}
}



\maketitle

\begin{abstract}
Traditional spam classification requires the end-users to reveal the content of their received emails to the spam classifier. On the other hand, spam classification over encrypted emails enables the classifier to classify spam email without accessing the email, hence protects the privacy of email content. In this paper, we construct a spam classification framework that enables the classification of encrypted emails. Our model is based on a neural network with a quadratic network component and a multi-layer perception network component. The quadratic network architecture is compatible with the operation of an existing quadratic functional encryption scheme that enables our classification to predict the label of encrypted emails. The evaluation results on real-world spam datasets indicate that our proposed spam classification model achieves accuracy over 96\%.

\end{abstract}


\section{Introduction}

Spam classification is one of the very early successful applications of machine learning in cybersecurity. Since the early work that proposed Naive Bayesian spam classifiers~\cite{sahami1998bayesian,rennie2000ifile}, spam email classification has evolved significantly over the years~\cite{taylor2007war}. As of now, some consider it as an almost solved problem~\cite{forbes,computerworld}. For example, Gmail claims to have reached over 99.9\% accuracy in 2015 by augmenting their rule-based spam filters with continually updating machine learning models implemented using Tensorflow~\cite{venturebeat,GoogleBlog,wired}.


Nonetheless, all of the existing solutions have the limitation of the end-users needing to reveal the email in plaint-text to the spam classifier (more generally the model owner), which is usually implemented as an in-built function of the email server. As such, in the current form the end-user cannot have both end-to-end encryption and spam classification together. In fact, this limitation is not only applicable for spam classification, but also for any other automated smart email classification tasks such as identifying urgent emails or separating personal and work emails. Therefore, it is important to come up with solutions that can conduct email classification without revealing the actual content to the email service provider.


Recently, machine learning on encrypted data has received wide attention because of the advances in \textit{fully homomorphic encryption (FHE)}. While there are FHE schemes proposed for spam classification~\cite{khedr2015shield,pathak2011privacy}, they have several practical limitations. First, FHE schemes require back and forth communications between the end-user and the model owner. This is because in FHE schemes, the model owner sends the encrypted model to the end-user who gives the plain-text message as the input to the encrypted model to get the encrypted prediction vector. Next, the end-user sends the encrypted prediction vector back to the model owner to get it decrypted and the result is sent back. The second limitation is the fact that the end-user or some designated client on behalf the end-user needs to be online to make this mechanism working. Finally, spam emails reach the end-user's mailbox and consume resources to make the prediction. An ideal set-up for an asynchronous application such as email is where a server is making the email classification and forwarding only required emails to the end-user. In contrast to FHE, the emerging area of \textit{functional encryption}~\cite{FE_Boneh}  can support such a setup.

In functional encryption, the public key $pk$ encrypts the plain-text $x$ to form a cipher-text $c = enc(pk, x)$. Then, secret key $sk$ is used to decrypt the cipher-text $c$ resulting in a new data $y = dec(sk, c)$ where $y$ meets a specific functionality according to the functional encryption scheme: $y = F(sk, x)$~\cite{reading_in_the_dark, fe_quadratic_function}. 
In this paper, we extend the work of  Ryffel et al.~\cite{RyffelPBDG19} to build a deep learning model for privacy-preserving text classification on encrypted data using functional encryption. Particularly, we build a neural network which consists of a quadratic component and a multi-layer perception component. The quadratic network is compatible to a quadratic functional encryption scheme to enable the model to make predictions on encrypted data. In particular, the input to the quadratic network is the encrypted data and the output of the quadratic network is a plain-text vector prediction which is then fed into the multi-layer perception network to obtain the final label of the encrypted data. 




We make the following contributions in this paper.

\begin{itemize}
    \item We propose a practical text processing pipeline that allows to integrate quadratic functional encryption into a neural network. The proposed pipeline enables spam classification at a server without decrypting emails, thus ensures user privacy. In contrast to FHE-based solutions, the proposed pipeline does not require back and forth communications between the end-user and the server.
    \item We implement a prototype of a spam filter system and evaluate its performance with multiple real world spam datasets. Our analysis show that there is no significant accuracy loss in the proposed pipeline compared to existing plain-text-based systems.
    

    \item We evaluate the performance of the spam filter system on functional encrypted data in terms of computational overhead and show that on average prediction time for our system is in the range of 22s--40s per email depending on the vocabulary size.

    \item Finally, we also compare  the performance of our system with FHE-based logistic regression classifier and show that both achieve similar levels of accuracy. 
 However, due to the simple nature of the model and the fact that in FHE-based scheme the model is encrypted than the data, we show that FHE-based scheme can do much faster predictions.
    
    

\end{itemize}

The rest of the paper is organised as follows. In Section~\ref{sec:background} we provide a brief background of functional encryption and homomorphic encryption and in Section~\ref{sec:relatedwork} we present related work. We describe our proposed privacy-preserving spam filter framework in Section~\ref{sec:spam_filter_framework} and present our results in Section~\ref{sec:performance}. We discuss limitations of our work and possible extensions in Section~\ref{sec:discussion}. Section~\ref{sec:conclusion} concludes the paper.

\section{Background}\label{sec:background}

In this section, we provide some background information about functional encryption and homomorphic encryption that is required to fully explain our spam classification pipeline and the baseline we use.

\subsection{Functional Encryption (FE)}

Functional encryption can be considered as a generalization of traditional public-key encryption. However, unlike the case of public-key encryption where the decryption key allows to obtain the plain-text back, the \textit{functional decryption key} allows to obtain only a specific functional evaluation of the plain-text~\cite{FE_Boneh}. That is, in a functional encryption scheme, given the cipher-text $c$ of a plain-text message $m$, the decryption key $dk_f$
corresponding to some function $f$, allows to obtain only $f(m)$ without revealing anything about $m$.



While the ideal scenario is to enable computation of an arbitrary function on encrypted data, as of now there are no functional encryption schemes for polynomials of an order greater than two~\cite{priv_enhance_fe}. Existing schemes support only linear functions and  quadratic functions. \\ \vspace{-3mm}


\noindent{{\bf a) Linear schemes}}, also called as inner-product schemes~\cite{simple_linear_fe, fe_linear_scheme} allow to obtain the inner product between the plain-text and a given arbitrary vector without decrypting the cipher-text. That is, given an encryption of a vector $ \boldsymbol{x} \in Z^n$ (denoted by $\boldsymbol{c}$) and a functional decryption key $dk_f$ generated over a vector of coefficients $\boldsymbol{y} \in Z^n$, one can obtain $f(\boldsymbol{x},\boldsymbol{y}) = \boldsymbol{x}.\boldsymbol{y}$, which is the inner product of vectors $\boldsymbol{x}$ and $\boldsymbol{y}$. Note that the result of this evaluation only reveals value of inner product and never reveals the underlying plain-text data. \\ \vspace{-3mm}




\noindent{{\bf b) Quadratic schemes}} are a new class of  functional encryption schemes~\cite{reading_in_the_dark, fe_quadratic_function} used to represent the degree-two multivariate quadratic polynomials over integers. They are more useful because linear functional encryption schemes can only compute simple statistics of encrypted data. For example, quadratic schemes can compute more complex functions such as weighted mean, variance, co-variance, and root-mean-square on encrypted data. 


More specifically, in quadratic schemes given the encryption $\boldsymbol{c}$ of vector $\boldsymbol{x} \in Z^n$ and a key $dk_Q$ associated with a square matrix $Q \in Z^{n \times n}$, one can decrypt $\boldsymbol{c}$ using $dk_Q$ to obtain $f(\boldsymbol{x}) = \boldsymbol{x}^T Q \boldsymbol{x}$, which is the quadratic-product of $\boldsymbol{x}$ and $Q$. 







\subsection{Homomorphic Encryption (HE)}
Similar to functional encryption, homomorphic encryption (HE)~\cite{fhe_first} also allows computations on encrypted data. However, here the result is in the encrypted form, which can only be decrypted by the owner of the secret key.


There are three main types of homomorphic encryption including partially homomorphic encryption (PHE), somewhat homomorphic encryption (SHE), and fully homomorphic encryption (FHE)~\cite{he_survey}. The primary differences among them are the types and frequency of mathematical operations that can be performed on cipher-text. Specifically, PHE allows only one type of operation with an unlimited number of times to be performed on encrypted data. Some examples of PHE include ElGamal encryption (allowing a multiplication on cipher-text), Elliptic Curve ElGamal and Paillier encryption (allowing an addition on cipher-text). SHE supports some types of operations up to a certain complexity with a limited number of times. FHE allows any operations with an unlimited number of times. 

While FHE is still nascent, practical PHE and SHE schemes do exist as open-source implementations. For example, Paillier encryption~\cite{PythonPaillier} is an additive HE scheme that allows addition and scalar multiplication operations on encrypted data. The Microsoft SEAL~\cite{sealcrypto}, is a homomorphic encryption library that allows additions and multiplications to be performed on encrypted integers or real numbers. 


Although HE ensures private computation on encrypted data, it has limitations in many practical applications where the result of the computation is needed for immediately actions such as spam filtering services. In fact, in HE, the computation on encrypted data results in an encrypted output which needs to be sent back to the owner of the secret key for decryption.




\section{Related Work}\label{sec:relatedwork}

In this section, we first summarize work in machine learning based spam classification over plain-text data. Next, we present related work on privacy-preserving classification in general using homomorphic encryption and functional encryption, as well as specific work that focused on spam classification.  

\subsection{Spam Classification using Plain-text} 

Spam classification is one of the early success stories of machine learning in cybersecurity. Over the last two decades machine learning has replaced knowledge engineering based traditional spam classification~\cite{history_of_spam}. Starting from the simplistic Naive Bayes classifier~\cite{sahami1998bayesian,rennie2000ifile} a range of other classifiers such as Support Vector Machines~\cite{svm}, Logistic Regression~\cite{LR}, K-Nearest Neighbor~\cite{knn_spam_class}, and Decision Trees~\cite{compre_survey_spam} have been proposed for spam classification.








While the machine learning based spam classification schemes have achieved nearly perfect accuracy on plain-text data, concerns about the privacy of the individuals who contribute their data for model training and the leakage of private information of the emails to the classifier at inference time, make a case for privacy-preserving classification. 

Thus, machine learning based spam classification in which the classifier has no direct access to plain-text data has recently received significant attention. In what follows, we present privacy-preserving classifications using homomorphic encryption (HE) and functional encryption (FE) in general and summarize specific works that focused on spam classification.

\subsection{Privacy Preserving Classification using HE} 
Homomorphic Encryption (HE) schemes allow computations on encrypted data.  Specifically, the model owner and the data owner can perform a classification task without revealing the model or the data to each other. Multiple work proposed privacy preserving classification using the advantages of HE~\cite{xu2019cryptonn,private_class_ndss15}.
For instance, in~\cite{private_class_ndss15} the authors assume that a pre-trained model is available and show how HE building blocks can be combined to enable private prediction.


In~\cite{khedr2015shield}, the authors proposed optimizations on top of a current homomorphic encryption implementation and constructed a secure email spam filter on top of it. The proposed spam filter is based on the Naive Bayes classifier~\cite{sahami1998bayesian} where the model of the spam filter is trained on plain-text emails. At inference time, the spam filter works on homomorphic encrypted emails.

Pathak et al.~\cite{pathak2011privacy} proposed protocols for a logistic regression based spam classification. The proposed protocols enable both private training and private prediction by combining the advantages of homomorphic encryption and secure multiparty computation. Particularly, in the training phase, the classification model is trained over datasets from multiple participants in a way that the participants compute gradients on their dataset and the model parameters are then updated by privately aggregating these gradients. In the prediction phase, the model is encrypted with homomorphic encryption and sent to the client who wishes to classify its emails. The encrypted result from the client is randomized before being sent back to the model owner who can obtain the plain-text of the randomized result. A variant of the secure millionaire protocol is executed by the client and the model owner to learn a result of a comparison which is then translated to the final result.

\subsection{Privacy Preserving Classification using FE} 
The conceptual idea of using functional encryption for private spam classification was proposed by Dufour-Sans~\cite{reading_in_the_dark} and Ryffel~\cite{RyffelPBDG19}. Here, the classification model is trained on plain-text data. Then, the trained model weights are used to generate a functional decryption key which allows the prediction on encrypted input data. The prediction process results in the label of the encrypted data in clear without revealing the underlying plain-text data. 

Ligier et al.~\cite{icissp17} proposed privacy preserving data classification using inner-product functional encryption scheme. The input data is encrypted by the functional encryption. This work enables the prediction on encrypted data by first extracting the inner-products and then producing the label of the encrypted input data. The proposed scheme is evaluated with the MNIST dataset for image classification. 

In~\cite{reading_in_the_dark}, the authors proposed a new functional encryption scheme for quadratic polynomials. This work defined quadratic networks and demonstrated that such network designs are suited for the functional encryption of quadratic functions.  In~\cite{RyffelPBDG19}, Ryffle et al. extended ~\cite{reading_in_the_dark} and proposed a partially encrypted neural network for image classification. One part of the neural network is composed of a quadratic network that is compatible with quadratic functional encryption schemes. The second part of the neural network takes the output of the quadratic network as the input, sends it through a few more fully connected layers and finally does the prediction. 


Although functional encryption schemes were introduced and demonstrated for image classification tasks, to the best of our knowledge our work is the first to develop a practical privacy preserving email spam classification framework using quadratic functional encryption.


\section{Privacy-Preserving Spam Filter Framework} \label{sec:spam_filter_framework}

In this section, we describe the proposed functional encryption-based spam classification framework. Our work is using the functional encryption (FE) scheme proposed by Dufour-Sans et al.~\cite{reading_in_the_dark} and the partially encrypted neural network architecture proposed by Ryffel et al.~\cite{RyffelPBDG19}. Based on these two building blocks, we propose a practical text processing pipeline that allows integrating a quadratic functional encryption scheme into a neural network that can be used for spam email classification at email server level.  


\subsection{Partially Encrypted Neural Network using Quadratic FE}

We classify text data which can be represented as a vector $\boldsymbol{x} \in Z^n$, where $n$ represents the feature vector length. We build a model $(q_i)_{i \in [l]}$ for each label $i \in [l]$, such that our prediction $\boldsymbol{y}$ for $\boldsymbol{x}$ is $\text{argmax}(q_i(\boldsymbol{x}))$. According to \cite{reading_in_the_dark}, the functional encryption scheme for quadratic polynomials can form a degree-two polynomial network with one hidden layer. Thus, we build our model as $q_i(\boldsymbol{x}) = (P\boldsymbol{x})^T D_i (P\boldsymbol{x})$, $\forall i \in [l]$ where a bias term is added to vector $\boldsymbol{x}$ so that $\boldsymbol{x} = (1 \ x_1 \ \dots \ x_n)$. Note that, $D_i \in Z^{d \times d}$ is a diagonal matrix which contains at most $d$ non-zeros entries, and $P \in Z^{n \times d}$ is used to project the input vector $\boldsymbol{x}$ from $Z^n$ to $Z^d$. Thus, our model is a quadratic network with one hidden layer of $d$ neurons.

\begin{figure}
    \centering
    \includegraphics[width=0.8\linewidth]{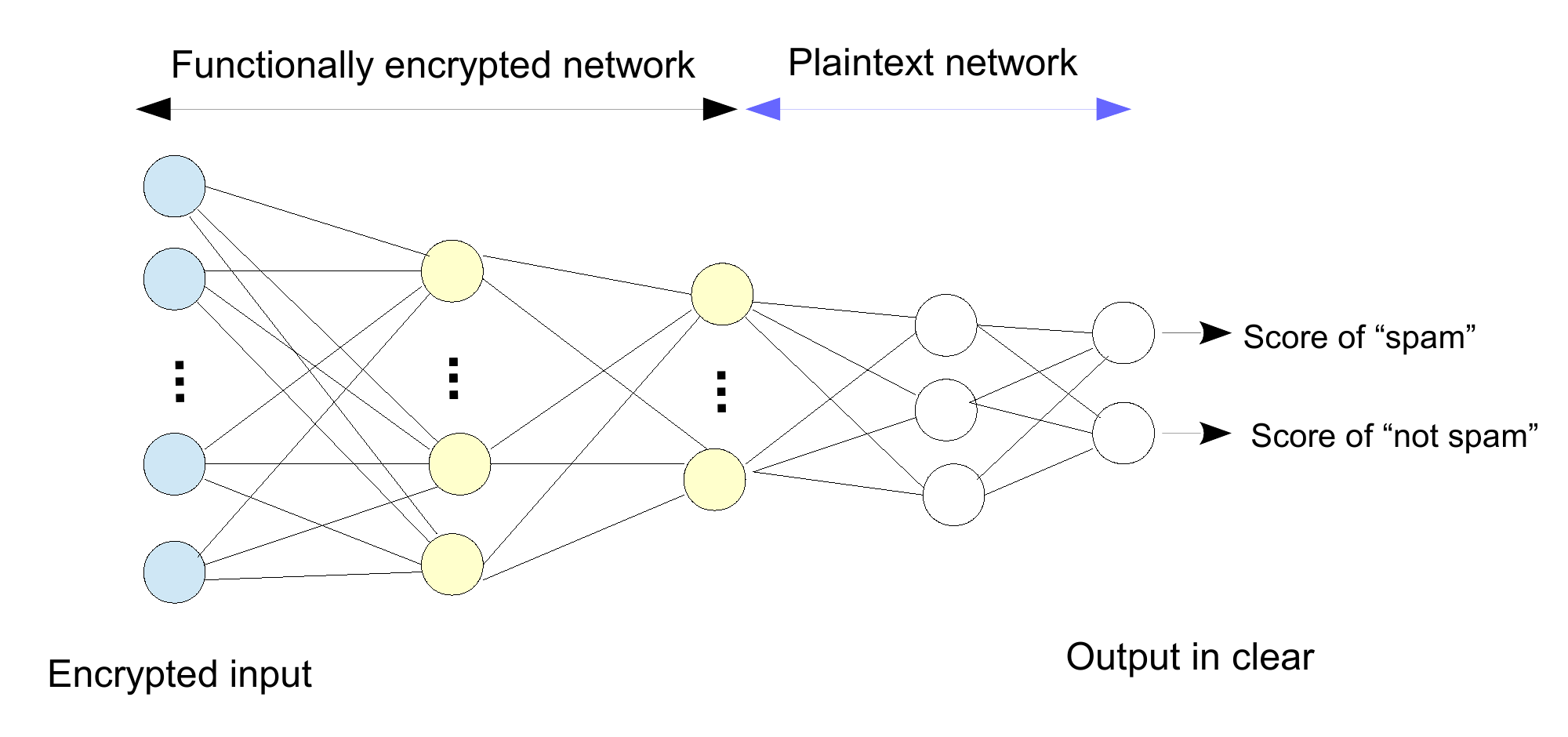}
    \caption{Partially encrypted network using quadratic FE scheme}
    \label{fig:network}
\end{figure}

Given an input vector $\boldsymbol{x}$, the output of the quadratic network is $( q_i(\boldsymbol{x}))_{i \in [l]}$ which is in plain-text. As discussed in~\cite{RyffelPBDG19}, adding more plain-text layers on top of a quadratic network could help improving the overall accuracy of the classification. Therefore, we build our model as illustrated in Figure~\ref{fig:network}. We show the model parameters in Table~\ref{tab:model_architecture}.  

\begin{table}[]
    \centering
    \caption{Model architecture}
    \begin{tabular}{c|c}\hline
        {\bf Encrypted Network} & {\bf Parameters}  \\
        \hline
         Dense & (n, 40)  \\
         \hline
         Lambda & Lambda $\boldsymbol{x}$ : $\boldsymbol{x}$*$\boldsymbol{x}$  \\
         \hline
         Dense & (40, 20)  \\
         \hline
         {\bf Plain-text Network} & \\
         \hline
         Dense & (20, 10)  \\
         \hline
         Dense & (10, 2), act = Sigmoid \\
         \hline
    \end{tabular}
    \label{tab:model_architecture}
\end{table}

Our model can be seen as a partially encrypted network which consists of two parts: \textit{an encrypted part} and \textit{a plain-text part}. The encrypted part of the network consists of an input layer of size $n$, and a hidden layer of size 40. Next is a quadratic function \textit{lambda} that does element-wise squaring, followed by a hidden layer of  size 20. The lambda layer is there to make sure that the encrypted part of the model is compatible with the quadratic FE scheme.


The plain-text part of the network takes the $(1 \times 20)$ output vector of the encrypted network as its input, then sends it through a dense layer of size 10.
The final layer is a size 2 dense layer with sigmoid activation that produces the final prediction. Though our plain-text network consists of only two layers, this part of the network can be extended as much as one would like to increase the model complexity to facilitate large and more complex datasets.

\subsection{Text Processing Pipeline}

The input to our neural network must be numerical feature vectors with a fixed size rather than the raw text documents of variable length. This makes integrating text data into a functional encryption pipeline less trivial compared to images. First, since functional encryption scheme uses the asymmetric groups to encrypt the data, the inputs must consist integers only. Therefore, we \textit{vectorize} emails using \textit{document-term-matrix} to convert the text data into integer vectors.



However, the size of the input vector directly impact the computational overhead of the encrypted network. This is a rather challenging problem in text classification since the input vector length is the vocabulary size. The large size of the input vector results in expensive computational overhead in the encrypted network. To mitigate this, we use \textit{information gain} to select the $n$ most discriminative features.



\subsection{Model Training}
\label{sec:training}

We train the model on plain-text data. The raw text data is turned into numerical feature vectors and then fed into our model. Our trained model includes weights and bias vectors for nodes in each layer. We then use the model parameters of layers in the encrypted network to generate a functional decryption key that allows to obtain the decrypted intermediate layer output of the encrypted input vector. The model parameters of layers in the plain-text network are used to predict the final output (i.e. the label of the encrypted input). \\ \vspace{-2mm} 

\noindent{\textbf{Post-training quantization:} We note that in order to be compatible with the FE encryption scheme (which uses bi-linear groups based on integers), the weight and bias of each node in the encrypted network part must be  converted into integers. To achieve this, we apply post-training quantization~\cite{tensorflow}}. 

\subsection{Threat Model}

The overall framework of the email spam filter includes sender's email client, sender's email server, recipient's email client and recipient's email server. We assume that the recipient has a pre-trained classification model and generates a functional decryption key $dk_f$ based on the pre-trained model and the master secret key. The recipient then provides this $dk_f$ and the pre-trained model to their email server.



In our threat model, we consider recipient's email server as an \textit{honest-but-curious} entity. That is, the server faithfully performs assigned operations but tries to infer additional information from what it receives. The recipient wants to use the server to separate out spam emails so that such emails do not come to their inbox, without revealing the actual content.



The overall flow of the scheme is as follows. The email sender encrypts the email using the recipient's public key $pk_R$, generates a document-term-matrix (DTM) vector of the text email, encrypts the vector using the same public key and sends both of these to the recipient's email server. Here we assume that the sender's email client is trustworthy. This is not an unrealistic assumption because the recipient's email server can accept emails only from a set of known clients.

Upon the reception of the encrypted email and the encrypted DTM vector, the recipient's email server can do the spam classification using the $dk_f$ and the model previously shared by the recipient. If the email is classified as spam, the mail server does not forward that email to the recipient's email client. While we demonstrate this for spam classification, this process is generic and can be applied to any text-based email classification task.



\section{Performance Evaluation}\label{sec:performance}
\subsection{Datasets and the Processing Environment}
We evaluate the performance of our approach on three large, publicly available datasets that are commonly used in spam classification research: {TREC07p}, {CEAS08-1}, and {ENRON}. 

TREC07p is a dataset from 2007 TREC Public Spam Corpus,\footnote{\url{https://plg.uwaterloo.ca/~gvcormac/treccorpus07/}} which was originally
constructed for the TREC spam filtering competitions. CEAS08-1 dataset is from the CEAS 2008 Live Spam Challenge Corpus.\footnote{\url{https://plg.uwaterloo.ca/~gvcormac/ceascorpus/}}
The ENRON dataset\footnote{\url{http://www2.aueb.gr/users/ion/data/enron-spam/}} contains emails from 150 users of the company Enron. We use \textit{enron-1} and \textit{enron-2} folders in our evaluation.  


As emails in three datasets are in \textit{html} format, we extract the text from both email's subject and email's body and save them in text format. We apply the standard text pre-processing steps such as changing characters to lower case, removing non-alphabet characters, removing punctuation, removing words longer than 15 characters, removing tokens with numbers only, and stemming. 
At the end, we keep emails that contain at least 10 words and no more than 100 words and discard the others. 
We include an upper-bound for the email size to ensure that our vocabulary size is within reasonable limits. As we show later, the encryption and decryption times of the  functional encryption is linearly proportional to the vocabulary size.




After the pre-processing steps, for TREC07p, we have 19,178 emails that includes 11,361 spam and 7,817 ham. For CEAS08-1, we have 40,524 emails including 29,919 and 10,506 emails labeled as spam and ham respectively. Finally, for ENRON, we have 6,748 emails which include 1,885 and 4,863 emails labeled as spam and ham respectively. We split each dataset in 70:30 ratio to obtain the training and test sets.





We ran all of our subsequent experiments in Amazon EC2 using a \textit{r5a.large} having 2.5GHz AMD EPYC 7000 series processor and 16GB of memory.

\subsection{The Baseline Scheme}\label{subsec:baseline}

We use the logistic regression-based spam filtering scheme on partially homomorphic encrypted text data~\cite{PythonPaillier}, which is an additively homomorphic encryption scheme as our baseline.

In this scheme, the model owner trains a logistic regression model on plain-text data. After that, the model owner generates a public-private key pairs under the Paillier encryption scheme and encrypts the model with the public key. 

The model is then shared with email clients (i.e., email recipients who want to have the spam filtering feature). Once receiving the encrypted model, the email recipient simply applies it on their own plain-text email which results in an encrypted output. The email recipient sends this result back to the model owner who decrypts it using the private key and sends the plain-text prediction back to the email recipient. In this way, the email recipient does not learn anything about the model weights, and the model owner does not learn anything about the email content. We refer to this baseline as the ``\textit{HE model}'' for the rest of the paper.

\subsection{Prediction Accuracy}

\begin{figure}[ht!]
\centering
\includegraphics[width=0.8\linewidth]{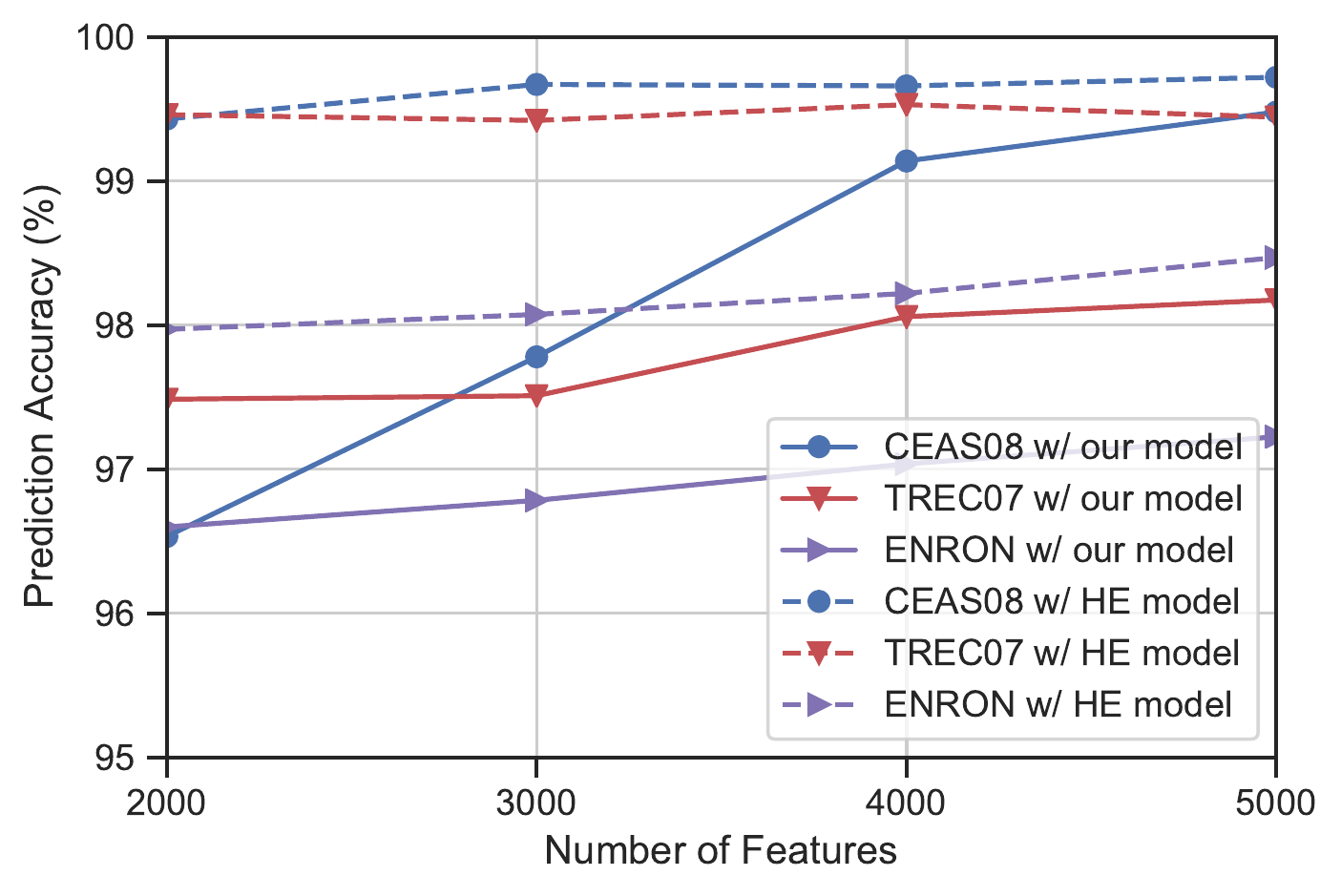}
\caption{Accuracy of the spam filter on encrypted data} 
\label{fig:prediction_acc}
\end{figure}

We evaluate the accuracy of our proposed spam filter on encrypted data by varying the number of features between 2,000 and 5,000. We use \textit{information gain} for feature selection. In Figure~\ref{fig:prediction_acc} we show the results for our model as well as the baseline HE model for all three datasets.

According to the figure while both the models have over 96\% accuracy, our model performance is slightly lower than the HE model, especially at lower vocabulary sizes. This performance gap is happening due to the quantization step (cf. Section~\ref{sec:training}). While this is a limitation, the true advantage of our scheme is that we avoid the back and forth communication between the server and the client (email recipient), and the server is in a  position to classify emails even when the client is not online. 








\subsection{Computational Overhead}

We next present the computational overhead results for the two schemes for both encryption and decryption steps. For clarity, we present results only for the ENRON dataset, and the other datasets follow similar patterns. 

\subsubsection{Encryption time} In our model, the encryption time involves taking the feature vector representing an email and encrypting it using the public key associated with a master secret key of the end user. This is a task that needs to be repeated for each email by the sender. 

In the baseline (HE model), the email sender also needs to take the same steps. There is also an additional one-off step to encrypt the model weights which is done by the model owner. As the model weights is a vector of the same size as the feature vector presenting an email, we observed that the encryption time of model weights vector and that of the feature vector of an email is similar. For clarity, we only show the runtime for encrypting the feature vector of an email in the \textit{HE model}. 

\begin{figure}[ht!]
\centering
\includegraphics[width=0.8\linewidth]{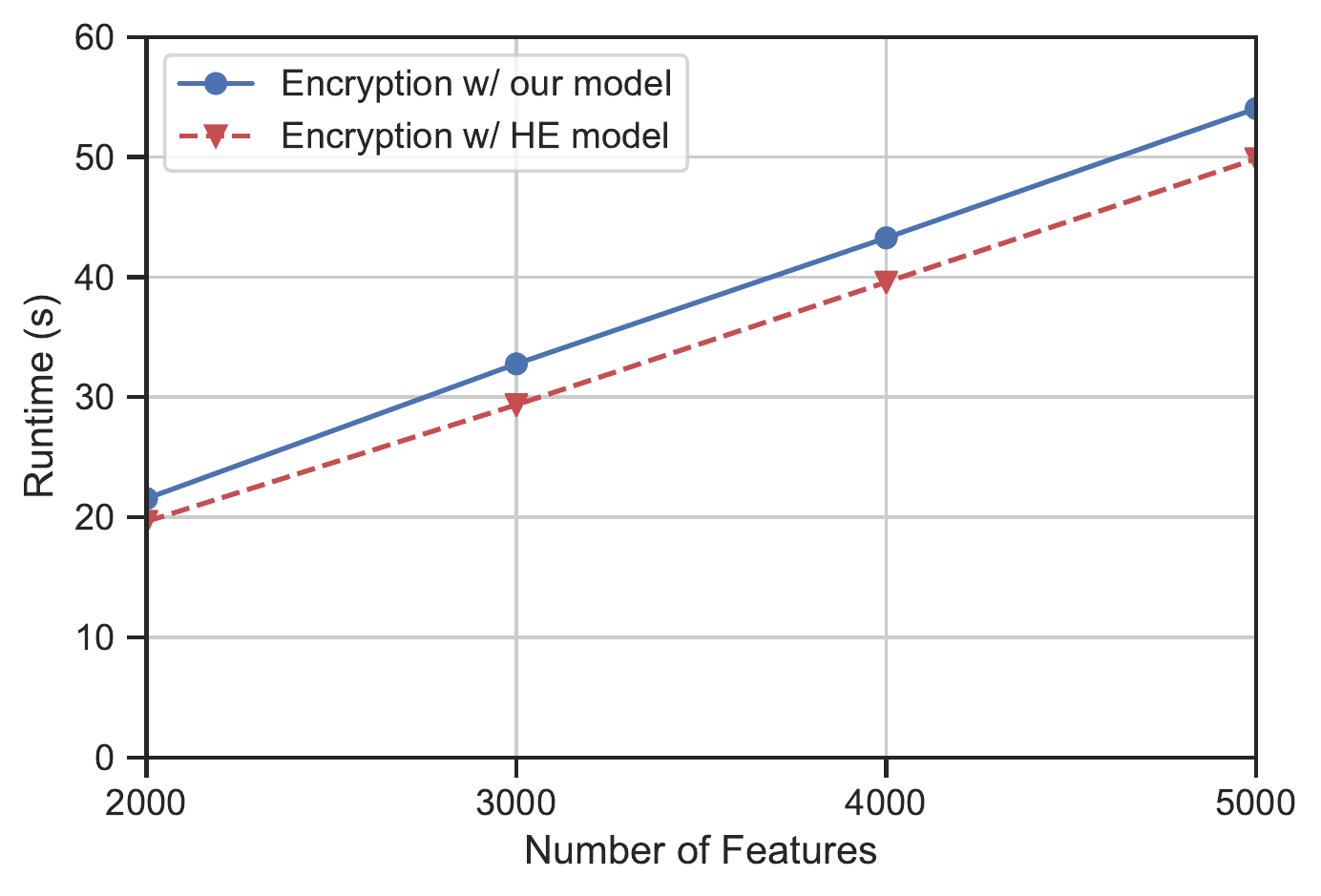}
\caption{Average encryption time for a single email \label{fig:fe_encryption}} 
\end{figure}

In Figure~\ref{fig:fe_encryption}, we show how encryption time varies with the feature vector size. According to the figure, we can see that the encryption time increases linearly with the feature vector size for both models. There is no significant difference between the times consumed by both the models. 



\subsubsection{Prediction time including decryption} 

Next, we compare the computation time of spam classification task in terms of \textit{average prediction time} per email for both the schemes. The prediction time in our scheme includes; \textit{i) the functional decryption time of the encrypted input vector} (i.e., obtaining the output of the encrypted part of the neural network) and \textit{ii) the classification time of the final label of the email} (i.e., get the final prediction from the plain-text part).

The total prediction time of the HE model consists of; \textit{i) the time taken by the client (i.e., the email recipient) to decrypt the email to get the plain-text} \textit{ii) the time taken by the client to get the encrypted prediction vector (i.e., multiplying the plain-text email input with the encrypted model weights)}, \textit{iii) the time taken to send the encrypted prediction to the server (model owner)}, \textit{iv) the time taken by the server to decrypt the encrypted prediction vector}, and \textit{v) the time taken to send the plain-text prediction vector back to the client}. Since we are doing the experiments in a single server, we highlight that times in iii) and v) are negligible in our experiments.

\begin{figure}[ht!]
\centering
\includegraphics[width=0.8\linewidth]{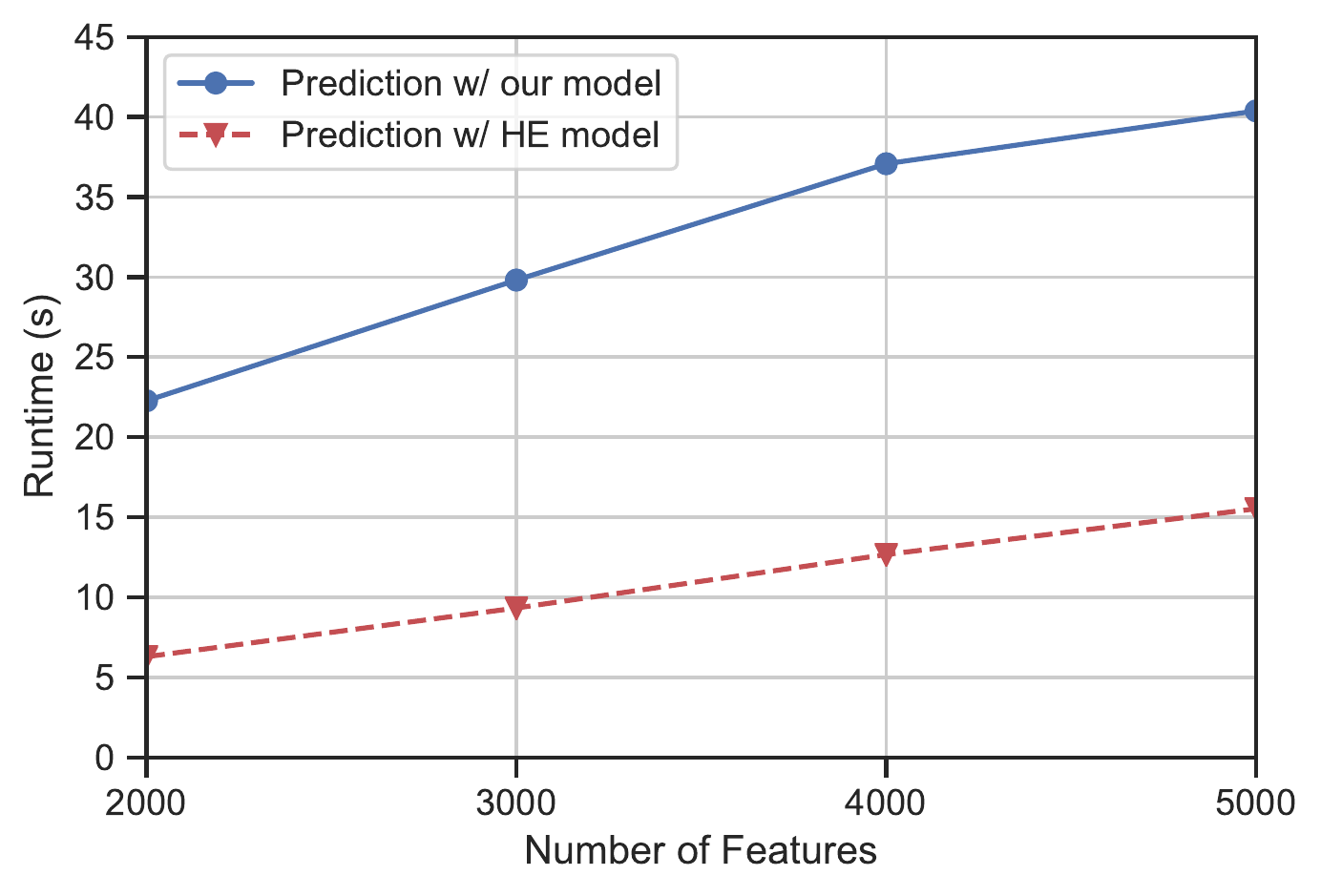}
\caption{Average prediction time for a single encrypted email \label{fig:fe_prediction}} 
\end{figure}



Figure \ref{fig:fe_prediction} shows that the average prediction time increases linearly with the size of the feature vector (the number of features used in the model). According to the figure, our model takes 20s--25s more time compared to the HE model. 

The functional decryption task of our model consists of two independent phases: the evaluation phase which covers the exponentiations and pairings to evaluate the underlying function and a discrete logarithm phase. 
The discrete logarithm step involves calculating discrete logarithms equivalent to the size of the output layer of the encrypted part of the model.
For example, for a feature vector size of 2,000, the decryption time is 25.87s which includes 8.97s for the evaluation phase and 16.90s for the discrete logarithm phase. We discuss possible ways of improving this time in Section~\ref{sec:discussion}. Also, as mentioned earlier, the current overhead calculation of the baseline HE model does not include the delay in the communications required between the email recipient and the model owner. 

\section{Discussion}
\label{sec:discussion}

In this section, we discuss the limitations of our work and possible future directions to follow to further improve the performance of the overall framework.

\subsection{Potential Information Leakage}
\label{ubsec:information_leak}

Our classification model is composed of an encrypted part and a plain-text part. It is possible that an adversary uses the output of the encrypted network and trains a new network to learn additional information. For example, in~\cite{RyffelPBDG19} the authors demonstrated this attack on an MNIST classifier and showed that the intermediate layer can leak information about the font of digits. In the case of our spam classification framework, the output of the encrypted part is a plain-text embeddings vector of $(1 \times t)$, ($t=20$ in our evaluation).  Using these embeddings, the adversary can try to learn the broader topic of the email or whether or not certain words are present in the email. 




We investigated this potential attack by choosing the top $k=10$ words with the highest information gain from the total vocabulary of the TREC07p dataset. We assigned a new \textit{private label} to each email based on the presence and absence of one of these words in an email. Then, we built a private learning model which takes the output vector of the encrypted part as its input and predicts the private label.





The evaluation result shows that the accuracy of the private learning model is over 90\% which implies that the adversary who has access to the output of the encrypted part of our spam classification could still learn the presence of some particular words in the original email. Consequently, if extra measures are not taken information about email content can be leaked at intermediate layers of the neural network. A possible solution to this issue is to apply \textit{collateral learning} as proposed by Ryffel et al.~\cite{RyffelPBDG19} for image classification to reduce the possible information leakage while maintaining the required level of accuracy for the spam classification task. We leave this direction as our future work.



\subsection{Computational Overhead Improvement} 
Although our model takes more time in the prediction task, there are further possible improvements. 
First, the evaluation phase in the decryption process evaluates the quadratic function on the encrypted input data. The computational overhead is increased with the vector length of the encrypted data. We can reduce this overhead by choosing a reasonable length of the encrypted vector which does not significantly reduce the prediction accuracy. As we can see from Figures~\ref{fig:prediction_acc} and~\ref{fig:fe_prediction}, choosing the vector length (number of features) of 4,000 instead of 5,000 can reduce the prediction time by 4s while not significantly dropping the prediction accuracy (only by 0.2\%--0.5\% for different spam datasets).  

Second, the discrete logarithm phase involves 20 discrete logarithms to retrieve an output vector of the encrypted part of the model. As also discussed in~\cite{reading_in_the_dark}, the discrete logarithm phase can be sped up by storing a large database of pre-computations. Specifically, the pre-computations allows the decryption task to avoid doing the discrete logarithm calculation online which sharply decreases the decryption time.


\subsection{Email Representation} 
In this work, we used the Document-Term-Matrix as the way of representing emails, whereas the state-of-the-art text classification work usually uses more advanced text representations based on pre-trained word vectors such as BERT embeddings~\cite{devlin2018bert}. Such embeddings are known to improve the performance of text classification because they are trained on large datasets and capture the context of the words unlike the Document-Term-Matrix where local context of the words is ignored. However, since the input to the encrypted network accepts only integer inputs, BERT-like embeddings cannot be readily integrated with the current pipeline. Generating quantized embeddings from transformer networks such as BERT is an active research topic~\cite{tissier2019near,grzegorczyk2016binary}, and can be attempted to further improve the accuracy.  

\subsection{Model quantization} 
We used Tensorflow in-built quantization to convert our model weights and biases to 8-bit integers. This step reduces the accuracy of the model. A possible improvement here is to quantize only the layers that are within the encrypted network or to use quantization-aware-training~\cite{jacob2018quantization}. Such steps are likely to reduce the drop in accuracy due to quantization.

\section{Conclusion}\label{sec:conclusion}
In this paper, we proposed a privacy-preserving spam classification framework using functional encryption. Our practical text processing pipeline allows the integration of quadratic functional encryption scheme into a neural network. Our experimental results indicate that the proposed classification model obtains high levels of accuracy of above 96\% for three real-world spam datasets. We also, compared our method with a homomorphic encryption-based scheme. While our method have high computational overheads compared to the HE scheme, our pipeline does not require back and forth communication between the sever and the email client, and as such more appropriate for server-based spam classification.

\bibliographystyle{IEEEtran}
\bibliography{IEEEabrv,refs.bib}

\begin{thebibliography}{10}
\providecommand{\url}[1]{#1}
\csname url@samestyle\endcsname
\providecommand{\newblock}{\relax}
\providecommand{\bibinfo}[2]{#2}
\providecommand{\BIBentrySTDinterwordspacing}{\spaceskip=0pt\relax}
\providecommand{\BIBentryALTinterwordstretchfactor}{4}
\providecommand{\BIBentryALTinterwordspacing}{\spaceskip=\fontdimen2\font plus
\BIBentryALTinterwordstretchfactor\fontdimen3\font minus
  \fontdimen4\font\relax}
\providecommand{\BIBforeignlanguage}[2]{{%
\expandafter\ifx\csname l@#1\endcsname\relax
\typeout{** WARNING: IEEEtran.bst: No hyphenation pattern has been}%
\typeout{** loaded for the language `#1'. Using the pattern for}%
\typeout{** the default language instead.}%
\else
\language=\csname l@#1\endcsname
\fi
#2}}
\providecommand{\BIBdecl}{\relax}
\BIBdecl

\bibitem{sahami1998bayesian}
M.~Sahami, S.~Dumais, D.~Heckerman, and E.~Horvitz, ``A bayesian approach to
  filtering junk e-mail,'' in \emph{Learning for Text Categorization: Papers
  from the 1998 workshop}, vol.~62, 1998, pp. 98--105.

\bibitem{rennie2000ifile}
J.~Rennie, ``ifile: An application of machine learning to e-mail filtering,''
  in \emph{Proc. KDD 2000 Workshop on Text Mining, Boston, MA}.

\bibitem{taylor2007war}
B.~Taylor, D.~Fingal, and D.~Aberdeen, ``The war against spam: A report from
  the front line,'' 2007.

\bibitem{forbes}
\BIBentryALTinterwordspacing
G.~Marks, ``How spam was solved,'' 2011. [Online]. Available:
  \url{https://www.forbes.com/sites/quickerbettertech/2011/10/17/how-spam-was-solved}
\BIBentrySTDinterwordspacing

\bibitem{computerworld}
\BIBentryALTinterwordspacing
M.~Elgan, ``Has the spam problem been solved?'' 2011. [Online]. Available:
  \url{https://www.computerworld.com/article/2498784/has-the-spam-problem-been-solved-.html}
\BIBentrySTDinterwordspacing

\bibitem{venturebeat}
\BIBentryALTinterwordspacing
K.~Wiggers, ``Gmail is now blocking 100 million more spam emails a day, thanks
  to tensorflow,'' 2019. [Online]. Available:
  \url{https://www.computerworld.com/article/2498784/has-the-spam-problem-been-solved-.html}
\BIBentrySTDinterwordspacing

\bibitem{GoogleBlog}
\BIBentryALTinterwordspacing
{Google {B}log}, ``How machine learning in g suite makes people more
  productive,'' 2017. [Online]. Available:
  \url{https://www.blog.google/products/g-suite/how-machine-learning-g-suite-makes-people-more-productive/}
\BIBentrySTDinterwordspacing

\bibitem{wired}
\BIBentryALTinterwordspacing
C.~Metz, ``Google says its ai catches 99.9 percent of gmail spam,'' 2015.
  [Online]. Available:
  \url{https://www.wired.com/2015/07/google-says-ai-catches-99-9-percent-gmail-spam/}
\BIBentrySTDinterwordspacing

\bibitem{khedr2015shield}
A.~Khedr, G.~Gulak, and V.~Vaikuntanathan, ``Shield: scalable homomorphic
  implementation of encrypted data-classifiers,'' \emph{IEEE Transactions on
  Computers}, vol.~65, no.~9, pp. 2848--2858, 2015.

\bibitem{pathak2011privacy}
M.~A. Pathak, M.~Sharifi, and B.~Raj, ``Privacy preserving spam filtering,''
  \emph{arXiv preprint arXiv:1102.4021}, 2011.

\bibitem{FE_Boneh}
D.~Boneh, A.~Sahai, and B.~Waters, ``Functional encryption: Definitions and
  challenges,'' in \emph{Theory of Cryptography}, 2011, pp. 253--273.

\bibitem{reading_in_the_dark}
E.~Dufour-Sans, R.~Gay, and D.~Pointcheval, ``Reading in the dark: Classifying
  encrypted digits with functional encryption,'' Cryptology ePrint Archive,
  Report 2018/206, 2018.

\bibitem{fe_quadratic_function}
C.~E.~Z. Baltico, D.~Catalano, D.~Fiore, and R.~Gay, ``Practical functional
  encryption for quadratic functions with applications to predicate
  encryption,'' Cryptology ePrint Archive, Report 2017/151, 2017.

\bibitem{RyffelPBDG19}
T.~Ryffel, D.~Pointcheval, F.~R. Bach, E.~Dufour{-}Sans, and R.~Gay,
  ``Partially encrypted deep learning using functional encryption,'' in
  \emph{AAnnual Conference on Neural Information Processing Systems}, 2019.

\bibitem{priv_enhance_fe}
T.~Marc, M.~Stopar, J.~Hartman, M.~Bizjak, and J.~Modic, ``Privacy-enhanced
  machine learning with functional encryption,'' in \emph{Computer Security --
  ESORICS 2019}, 2019.

\bibitem{simple_linear_fe}
M.~Abdalla, F.~Bourse, A.~De~Caro, and D.~Pointcheval, ``Simple functional
  encryption schemes for inner products,'' in \emph{Public-Key Cryptography --
  PKC 2015}, J.~Katz, Ed., 2015.

\bibitem{fe_linear_scheme}
S.~Agrawal, B.~Libert, and D.~Stehl{\'e}, ``Fully secure functional encryption
  for inner products, from standard assumptions,'' in \emph{Advances in
  Cryptology -- CRYPTO 2016}, 2016.

\bibitem{fhe_first}
C.~Gentry, ``A fully homomorphic encryption scheme,'' Ph.D. dissertation,
  Stanford, CA, USA, 2009.

\bibitem{he_survey}
A.~Acar, H.~Aksu, A.~S. Uluagac, and M.~Conti, ``A survey on homomorphic
  encryption schemes: Theory and implementation,'' \emph{ACM Comput. Surv.},
  vol.~51, no.~4, Jul. 2018.

\bibitem{PythonPaillier}
``Python paillier library,'' \url{https://github.com/data61/python-paillier}.

\bibitem{sealcrypto}
``{M}icrosoft {SEAL} (release 3.6),'' \url{https://github.com/Microsoft/SEAL},
  Nov. 2020, microsoft Research, Redmond, WA.

\bibitem{history_of_spam}
\BIBentryALTinterwordspacing
E.~Ferrara, ``The history of digital spam,'' \emph{Commun. ACM}, vol.~62,
  no.~8, p. 82–91, Jul. 2019. [Online]. Available:
  \url{https://doi.org/10.1145/3299768}
\BIBentrySTDinterwordspacing

\bibitem{svm}
S.~S. Roy, A.~Sinha, R.~Roy, C.~Barna, and P.~Samui, ``Spam email detection
  using deep support vector machine, support vector machine and artificial
  neural network,'' in \emph{Soft Computing Applications}, 2018.

\bibitem{LR}
K.~{Pawar} and M.~{Patil}, ``Pattern classification under attack on spam
  filtering,'' in \emph{2015 IEEE International Conference on Research in
  Computational Intelligence and Communication Networks}, 2015.

\bibitem{knn_spam_class}
A.~{Sharma} and A.~{Suryawanshi}, ``{A Novel Method for Detecting Spam Email
  using KNN Classification with Spearman Correlation as Distance Measure},''
  \emph{International Journal of Computer Applications}, Feb. 2016.

\bibitem{compre_survey_spam}
A.~{Karim}, S.~{Azam}, B.~{Shanmugam}, K.~{Kannoorpatti}, and M.~{Alazab}, ``A
  comprehensive survey for intelligent spam email detection,'' \emph{IEEE
  Access}, vol.~7, pp. 168\,261--168\,295, 2019.

\bibitem{xu2019cryptonn}
R.~Xu, J.~B.~D. Joshi, and C.~Li, ``Cryptonn: Training neural networks over
  encrypted data,'' 2019.

\bibitem{private_class_ndss15}
R.~Bost, R.~A. Popa, S.~Tu, and S.~Goldwasser, ``Machine learning
  classification over encrypted data,'' in \emph{22nd Annual Network and
  Distributed System Security Symposium, {NDSS}}, 2015.

\bibitem{icissp17}
D.~Ligier, S.~Carpov, C.~Fontaine, and R.~Sirdey, ``Privacy preserving data
  classification using inner-product functional encryption,'' in
  \emph{Proceedings of the 3rd {I}{C}{I}{S}{S}{P}}, 2017.

\bibitem{tensorflow}
\BIBentryALTinterwordspacing
``Post-training quantization,'' 2020. [Online]. Available:
  \url{https://www.tensorflow.org/lite/performance/post\_training\_quantization}
\BIBentrySTDinterwordspacing

\bibitem{devlin2018bert}
J.~Devlin, M.-W. Chang, K.~Lee, and K.~Toutanova, ``Bert: Pre-training of deep
  bidirectional transformers for language understanding,'' \emph{arXiv preprint
  arXiv:1810.04805}, 2018.

\bibitem{tissier2019near}
J.~Tissier, C.~Gravier, and A.~Habrard, ``Near-lossless binarization of word
  embeddings,'' in \emph{Proc. of the AAAI}, 2019.

\bibitem{grzegorczyk2016binary}
K.~Grzegorczyk and M.~Kurdziel, ``Binary paragraph vectors,'' \emph{arXiv
  preprint arXiv:1611.01116}, 2016.

\bibitem{jacob2018quantization}
B.~Jacob, S.~Kligys, B.~Chen, M.~Zhu, M.~Tang, A.~Howard, H.~Adam, and
  D.~Kalenichenko, ``Quantization and training of neural networks for efficient
  integer-arithmetic-only inference,'' in \emph{Proceedings of the IEEE
  Conference on Computer Vision and Pattern Recognition}, 2018.

\end{thebibliography}


\end{document}